\documentclass[aps,pra,twocolumn,groupedaddress]{revtex4-2}
\usepackage[dvips]{graphicx}
\usepackage[usenames,svgnames]{xcolor}
\usepackage{color}
\usepackage{url}
\usepackage{natbib}
\usepackage[colorlinks=true,linkcolor=magenta,citecolor=magenta,breaklinks=true]{hyperref}
\usepackage{amsmath,amssymb,bm}

\begin{document}

% Use the \preprint command to place your local institutional report
% number in the upper righthand corner of the title page in preprint mode.
% Multiple \preprint commands are allowed.
% Use the 'preprintnumbers' class option to override journal defaults
% to display numbers if necessary
%\preprint{}

%Title of paper
\title{Reducing errors and gate operations in digitized quantum annealing \\ with local counterdiabatic driving}

% repeat the \author .. \affiliation  etc. as needed
% \email, \thanks, \homepage, \altaffiliation all apply to the current
% author. Explanatory text should go in the []'s, actual e-mail
% address or url should go in the {}'s for \email and \homepage.
% Please use the appropriate macro foreach each type of information

% \affiliation command applies to all authors since the last
% \affiliation command. The \affiliation command should follow the
% other information
% \affiliation can be followed by \email, \homepage, \thanks as well.
\author{Takuya Hatomura}
\email[]{takuya.hatomura@ntt.com}
%\homepage[]{Your web page}
%\thanks{}
%\altaffiliation{}
\affiliation{Basic Research Laboratories \& Research Center for Theoretical Quantum Information, NTT, Inc., Kanagawa 243-0198, Japan}

%Collaboration name if desired (requires use of superscriptaddress
%option in \documentclass). \noaffiliation is required (may also be
%used with the \author command).
%\collaboration can be followed by \email, \homepage, \thanks as well.
%\collaboration{}
%\noaffiliation

\date{\today}

\begin{abstract}
Local counterdiabatic driving is a method of improving the performance of adiabatic control and digital implementation of quantum annealing with local counterdiabatic driving has been discussed. 
In this paper, we propose a decomposition formula which enables us to reduce digitization errors and the number of gate operations in digitized quantum annealing with local counterdiabatic driving. 
\end{abstract}

% insert suggested PACS numbers in braces on next line
\pacs{}
% insert suggested keywords - APS authors don't need to do this
%\keywords{}

%\maketitle must follow title, authors, abstract, \pacs, and \keywords
\maketitle

\section{\label{Sec.intro}Introduction}

Combinatorial optimization is a certain class of mathematical optimization problems~\cite{Korte2018}. 
Its applications cover a wide range of topics including practical issues, e.g., transportation~\cite{Neukart2017}, logistics~\cite{Rieffel2015}, finance~\cite{Rosenberg2016}, etc. 
However, the computational complexity of combinatorial optimization is generally NP-hard~\cite{Korte2018}, which requires an exponential time to be solved exactly. 
Therefore, it is of utmost importance to develop heuristic algorithms which enable us to obtain good approximate solutions within a polynomial time.

Quantum annealing was proposed as a method of obtaining the groud state or low-energy eigenstates of the Ising spin glass~\cite{Kadowaki1998}. 
We try to find the target ground state by adiabatically transforming a Hamiltonian from trivial one whose ground state is known to the target Ising spin glass. 
Quantum annealing is regarded as a heuristic quantum adiabatic algorithm for combinatorial optimization~\cite{Farhi2000,Hauke2020} because many combinatorial optimization problems can be formulated as ground-state search of the Ising spin glass~\cite{Lucas2014}. 
Performance of quantum annealing depends on adiabaticity of a process. 
For finding the exact solution or good approximate solutions, a long operation time is requried because of the adiabatic condition~\cite{Kato1950,Jansen2007}. 
In particular, an exponential time is necessary to overcome the first-order transition with an exponentially small energy gap.

Applications of additional fluctuations or biases, which are known as catalyst terms, have been discussed as means of improving quantum annealing~\cite{Seki2012,Albash2019,Grass2019,Albash2021}. 
In particular, antiferromagnetic $XX$-interaction terms introduce nonstoquasticity, which causes the sign problem in quantum Monte Carlo simulation, and thus antiferromagnetic $XX$-interaction terms are regarded as suitable catalysts introducing large quantum fluctuations. 
Removal of the first-order transitions in infinite-range (mean-field--like) models~\cite{Seki2012,Albash2019} and improvement of adiabaticity in a locally-interacting model~\cite{Albash2019} were reported. 
Introduction of bias $Z$-field terms was also considered~\cite{Grass2019,Albash2021}. 
Appropriate choice of biases improves performance of quantum annealing.

Shortcuts to adiabaticity~\cite{Torrontegui2013,Guery-Odelin2019,Hatomura2024} are also candidates for improving quantum annealing. 
Counterdiabatic driving of shortcuts to adiabaticity introduces additional driving (the counterdiabatic term) which completely cancels out nonadiabatic transitions~\cite{Demirplak2003,Demirplak2008,Berry2009}. 
Thus, the counterdiabatic term could be regarded as a well-designed catalyst. 
Exact application of counterdiabatic driving to quantum annealing is difficult because theoretical construction requires an exponentially large computational cost and experimental realization requires time-dependent control of many-body and nonlocal interactions. 
Variational and algebraic approaches enable us to approximately construct local counterdiabatic terms with ansatzes on operator forms~\cite{Sels2017,Hatomura2021,Takahashi2024,Bhattacharjee2023}. 
A detailed analysis of the counterdiabatic term revealed that $Y$-field terms and $YZ$-interaction terms are suitable for local counterdiabatic terms improving quantum annealing~\cite{Claeys2019}. 
It should be mentioned that a mean-field (classical) analysis of counterdiabatic driving applied to quantum annealing also suggests use of $Y$-field terms~\cite{Hatomura2017,Hatomura2018b}.

Quantum annealers which implement quantum annealing with thousands of qubits were already manufactured~\cite{King2018,King2022,King2023}. 
However, implementation of quantum annealing with catalysts is still challenging. 
Recently, gate-based quantum processors which enable us to implement universal quantum operations were developed~\cite{Arute2019,Wu2021,Bluvstein2024}. 
These gate-based quantum processors have potential to implement quantum annealing with catalysts. 
Indeed, digital realization of adiabatic state preparation~\cite{Steffen2003,Barends2016} and quantum annealing with local counterdiabatic driving ($Y$-field and $YZ$-interaction catalysts)~\cite{Hegade2021,Hegade2022} has been reported. 
Moreover, a gate-based algorithm for adiabatic transformation, which was inspired by counterdiabatic driving, was also proposed~\cite{Vreumingen2024} and its performance was numerically benchmarked~\cite{Hatomura2025a}.

Digitization of time-dependent Hamiltonian dynamics causes digitization errors~\cite{Huyghebaert1990,Lloyd1996}. 
Although intrinsic scaling advantage of digitized adiabatic state preparation~\cite{Kovalsky2023} and digitized counterdiabatic driving~\cite{Hatomura2022,Hatomura2023} was pointed out, large number of time slices is required for realizing target dynamics with high fidelity. 
In particular, adiabatic state preparation requires a long operation time for adiabaticity, and thus number of required time slices for adiabatic state preparation is much larger than that for digitization of general time-dependent Hamiltonian dynamics~\cite{Mbeng2019}. 
Suppression of these errors is of great importance for practical usefulness.

In this paper, we propose a decomposition formula which enables us to reduce digitization errors and the number of gate operations in digitize quantum annealing with local counterdiabatic driving. 
A key idea is the use of phase degrees of freedom in the measurement basis of interest, that is, we consider dynamics with the same measurement outcome as target dynamics. 
We use a property of unitary operators to digitize this dynamics and show that the present method is efficient compared with straightforward applications of the Suzuki-Trotter formulae.

%
%========================================================================
%
\section{\label{Sec.background}Background}

In this section, we give brief summaries of quantum annealing and Trotterization of time-dependent Hamiltonian dynamics.

%
%-------------------------------------------------------------------------
%
\subsection{Quantum annealing}

Quantum annealing is conducted with the following Hamiltonian
\begin{equation}
\hat{H}_\mathrm{QA}(t)=\lambda(t)\hat{H}_P+\bm{(}1-\lambda(t)\bm{)}\hat{V}
\label{Eq.ham.QA}
\end{equation}
where $\hat{H}_P$ is the problem Hamiltonian whose ground state is the solution of combinatorial optimization and $\hat{V}$ is the driver Hamiltonian whose ground state can easily be prepared~\cite{Kadowaki1998,Farhi2000}. 
The solution is obtained by adiabatically changing the time-dependent parameter $\lambda(t)$ from $\lambda(0)=0$ to $\lambda(T)=1$ with the annealing time $T$, which is large enough to satisfy the adiabatic condition~\cite{Kato1950,Jansen2007}.

Typically, the problem Hamiltonian is given by the Ising spin glass
\begin{equation}
\hat{H}_P=-\sum_{\substack{i,j=1 \\ (i<j)}}^NJ_{ij}\hat{Z}_i\hat{Z}_j-\sum_{i=1}^Nh_i\hat{Z}_i,
\label{Eq.probham}
\end{equation}
where $J_{ij}$ and $h_i$ are appropriate coupling strength and field strength, respectively. 
The driver Hamiltonian is usually given by the transverse-field Hamiltonian
\begin{equation}
\hat{V}=-\sum_{i=1}^N\Gamma_i\hat{X}_i,
\label{Eq.drivham}
\end{equation}
where $\Gamma_i$ is field strength. 
Here, the Pauli matrices of $N$ qubits are expressed as $\{\hat{X}_i,\hat{Y}_i,\hat{Z}_i\}_{i=1}^N$.

The introduction of catalysts to the quantum-annealing Hamiltonian (\ref{Eq.ham.QA}) is a means of improving quantum annealing. 
For example, in local counterdiabatic driving, we introduce the Pauli-Y transverse-field Hamiltonian
\begin{equation}
\hat{H}_\mathrm{LCD}(t)=\sum_{i=1}^N\alpha_i(t)\hat{Y}_i,
\label{Eq.ham.LCD}
\end{equation}
where $\alpha_i(t)$ is a time-dependent field strength. 
Its time dependence can be scheduled in various ways, e.g., the variational approach, the classical (mean-field) approach, etc. 
%In this paper, we adopt the variational approach and it gives
%\begin{equation}
%\alpha_i(t)=-\frac{1}{2}\frac{\dot{\lambda}(t)h_i\Gamma_i}{\lambda^2(t)h_i^2+\bm{(}1-\lambda(t)\bm{)}^2\Gamma_i^2+\lambda^2(t)\sum_{\substack{j=1 \\ (j\neq i)}}^NJ_{ij}^2}. 
%\end{equation}

%
%--------------------------------------------------------------------------
%
\subsection{\label{Sec.Trotterization}Digital quantum simulation using Trotterization}

We consider a time-evolution operator
\begin{equation}
\hat{U}(T,0)=\mathcal{T}\exp\left(-\frac{i}{\hbar}\int_0^Tdt\ \hat{H}(t)\right),
\label{Eq.tevo}
\end{equation}
with a time-dependent Hamiltonian $\hat{H}(t)$ and the operation time $T$. 
Note that $\mathcal{T}$ is the time-ordering operator. 
We introduce the discretized time $\{t_m\}_{m=0}^M$, where $0=t_0<t_1<t_2<\dots<t_M=T$ with an integer $M$, and then the time-evolution operator (\ref{Eq.tevo}) can be rewritten as
\begin{equation}
\begin{aligned}
\hat{U}(T,0)&=\prod_{m=M}^1\hat{U}(t_m,t_{m-1})\\
&=\prod_{m=M}^1\exp\left(\sum_{n=1}^\infty\hat{\Omega}_n^{(m)}\right),
\end{aligned}
\label{Eq.Magnus}
\end{equation}
where we adopt the Magnus expansion in the second line (see, e.g., Ref.~\cite{Ikeda2023}). 
The expanded terms are given by
\begin{equation}
\begin{aligned}
&\hat{\Omega}_1^{(m)}=-\frac{i}{\hbar}\int_{t_{m-1}}^{t_m}ds_1\ \hat{H}(s_1), \\
&\hat{\Omega}_2^{(m)}=-\frac{1}{2\hbar^2}\int_{t_{m-1}}^{t_m}ds_1\int_{t_{m-1}}^{s_1}ds_2\ [\hat{H}(s_1),\hat{H}(s_2)], 
\end{aligned}
\end{equation}
and so on. 
We can show that $\hat{\Omega}_n^{(m)}=\mathcal{O}(\delta t_m^{n+1})$ with the time step $\delta t_m=t_m-t_{m-1}$ for $n\ge2$ [the exception is that $\hat{\Omega}_1^{(m)}=\mathcal{O}(\delta t_m)$], and thus the time-evolution operator for each time slice can be simplified as
\begin{equation}
\hat{U}(t_m,t_{m-1})=\exp(\hat{\Omega}_1^{(m)})+\mathcal{O}(\delta t_m^3). 
\label{Eq.Magnus.approx}
\end{equation}

Then, we consider the decomposition of the time-evolution operator for each time slice (\ref{Eq.Magnus.approx}) into quantum gate operations. 
Suppose that the Hamiltonian is given by the following form
\begin{equation}
\hat{H}(t)=\sum_{k=1}^Ka_k(t)\hat{O}_k, 
\end{equation}
with implementable operators $\{\hat{O}_k\}_{k=1}^K$ and time-dependent coefficients $\{a_k(t)\}_{k=1}^K$. 
The simplest way of decomposition is the use of the Lie-Trotter formula (the first-order Suzuki-Trotter formula)
\begin{equation}
\exp(\hat{\Omega}_1^{(m)})=\prod_{k=1}^K\exp\left(-\frac{i}{\hbar}\int_{t_{m-1}}^{t_m}ds\ a_k(s)\hat{O}_k\right)+\mathcal{O}(\delta t_m^2),
\end{equation}
but it makes the scaling of errors worse. 
In contrast, the second-order Suzuki-Trotter formula gives
\begin{equation}
\begin{aligned}
\exp(\hat{\Omega}_1^{(m)})=&\left[\prod_{k=1}^K\exp\left(-\frac{i}{2\hbar}\int_{t_{m-1}}^{t_m}ds\ a_k(s)\hat{O}_k\right)\right]\\
&\times\left[\prod_{k=K}^1\exp\left(-\frac{i}{2\hbar}\int_{t_{m-1}}^{t_m}ds\ a_k(s)\hat{O}_k\right)\right]\\
&+\mathcal{O}(\delta t_m^3), 
\end{aligned}
\label{Eq.Suzuki2}
\end{equation}
that is, the order of errors is identical to that in Eq.~(\ref{Eq.Magnus.approx}). 
Note that decomposition formulae with better error scaling have also been proposed, but such decompositions require many quantum gate operations and the realization of such decompositions are not reasonable in current technology.

%
%----------------------------------------------------------
%
\subsection{Experimental realization}

The experimental realization of digitized quantum annealing was first reported in Ref.~\cite{Steffen2003}. 
In their experiment, they considered the second-order Suzuki-Trotter decomposition of a discretized time-evolution operator $\exp[-(i/\hbar)\delta t_m\hat{H}(t_m)]=\hat{U}(t_m,t_{m-1})+\mathcal{O}(\delta t_m^2)$ and used a nuclear magnetic resonance computer with three qubits. 
In Ref.~\cite{Barends2016}, the first-order Suzuki-Trotter decomposition of the discretized time-evolution operator was considered and a superconducting quantum computer with nine qubits was used, while the advantage of adopting Eq.~(\ref{Eq.Magnus.approx}) was mentioned. 
Notably, $XX$-interaction terms can also be realizable in their system. 
In Ref.~\cite{Hegade2021,Hegade2022}, the experimental realization of digitized quantum annealing with local counterdiabatic driving was reported, where the first-order Suzuki-Trotter decomposition of the discretized time-evolution operator was considered. 
In addition to $Y$-field terms, they also realized $YZ$-interaction terms.

%
%=======================================================================
%
\section{\label{Sec.method}Method}

%
%----------------------------------------------------------------------
%
\subsection{Decomposition formula}

We introduce a decomposition formula utilizing phase degrees of freedom which do not affect measurement outcomes. 
Suppose that $|\Psi(t)\rangle$ is target dynamics with a Hamiltonian $\hat{H}(t)$ and $\{|\sigma\rangle\}$ is a measurement basis of interest. 
Another dynamics $|\Psi_f(t)\rangle$ gives the same measurement outcome $|\langle\sigma|\Psi_f(t)\rangle|^2=|\langle\sigma|\Psi(t)\rangle|^2$ when it is given by
\begin{equation}
|\Psi_f(t)\rangle=\hat{V}_f(t)|\Psi(t)\rangle,
\label{Eq.state.phi}
\end{equation}
where $\hat{V}_f(t)$ is a unitary operator
\begin{equation}
\hat{V}_f(t)=\exp\left(i\sum_\sigma f_\sigma(t)|\sigma\rangle\langle\sigma|\right),
\label{Eq.Vf}
\end{equation}
with arbitrary phase $f_\sigma(t)$. 
By inversely solving the Schr\"odinger equation, we find that the Hamiltonian of the state (\ref{Eq.state.phi}) is given by
\begin{equation}
\hat{H}_f(t)=\hat{V}_f(t)\hat{H}(t)\hat{V}_f^\dag(t)+i\hbar\bm{(}\partial_t\hat{V}_f(t)\bm{)}\hat{V}_f^\dag(t). 
\label{Eq.ham.phi}
\end{equation}
We consider digitization of this dynamics (\ref{Eq.state.phi}) with the Hamiltonian (\ref{Eq.ham.phi}) instead of the original target dynamics $|\Psi(t)\rangle$ with the Hamiltonian $\hat{H}(t)$ since the measurement outcome is identical. 
This kind of the idea can also be found in another literature~\cite{Hatomura2023a,Hatomura2025}.

Now, we consider digitization of the time-evolution operator
\begin{equation}
\hat{U}_f(T,0)=\mathcal{T}\exp\left(-\frac{i}{\hbar}\int_0^Tdt\ \hat{H}_f(t)\right). 
\end{equation}
In addition to the Magnus expansion (\ref{Eq.Magnus}) with the approximation (\ref{Eq.Magnus.approx}), we consider the Taylor expansion of the Hamiltonian (\ref{Eq.ham.phi}) at the middle point, i.e., $\hat{H}_f(t)=\hat{H}_f(\tau_m)+\hat{H}_f^\prime(\tau_m)[t-\tau_m]+\mathcal{O}(\delta t_m^2)$ ($\tau_m=t_{m-1}+\delta t_m/2$) for $\hat{U}_f(t_m,t_{m-1})$. 
Then, we obtain
\begin{equation}
\hat{U}_f(t_m,t_{m-1})=\exp\left(-\frac{i}{\hbar}\delta t_m\hat{H}_f(\tau_m)\right)+\mathcal{O}(\delta t_m^3),
\label{Eq.Uf.Magnus}
\end{equation}
and it can be rewritten as
\begin{equation}
\begin{aligned}
&\exp\left(-\frac{i}{\hbar}\delta t_m\hat{H}_f(\tau_m)\right)=\\
&\hat{V}_f(\tau_m)\exp\left(-\frac{i}{\hbar}\delta t_m\hat{H}_f^V(\tau_m)\right)\hat{V}_f^\dag(\tau_m),
\end{aligned}
\label{Eq.decom}
\end{equation}
with
\begin{equation}
\begin{aligned}
\hat{H}_f^V(\tau_m)&=\hat{V}_f^\dag(\tau_m)\hat{H}_f(\tau_m)\hat{V}_f(\tau_m)\\
&=\hat{H}(\tau_m)+i\hbar\hat{V}_f^\dag(\tau_m)\bm{(}\partial_t\hat{V}_f(\tau_m)\bm{)}, 
\end{aligned}
\label{Eq.ham.phi.u}
\end{equation}
where we use a property of unitary transformation, i.e., $\exp(\hat{H})=\hat{V}\exp(\hat{V}^\dag\hat{H}\hat{V})\hat{V}^\dag$ for an operator $\hat{H}$ and a unitary operator $\hat{V}$. 
Remarkably, the formula (\ref{Eq.decom}) does not cause any additional error to Eq.~(\ref{Eq.Uf.Magnus}).

\subsection{Application to quantum annealing}

Now we discuss application of the decomposition formula (\ref{Eq.decom}) to quantum annealing. 
We can incorporate the enhancement with local counterdiabatic driving into this application, i.e., the total Hamiltonian can be
\begin{equation}
\hat{H}(t)=\hat{H}_\mathrm{QA}(t)+\hat{H}_\mathrm{LCD}(t). 
\end{equation}
Hereafter, we set $\hbar=1$ according to the conventional notation of quantum annealing.

In quantum annealing, we measure the final state $|\Psi(T)\rangle$ in the computational basis $|\sigma\rangle=|\sigma_1,\sigma_2,\dots,\sigma_N\rangle$, where $\hat{Z}_i|\sigma\rangle=\sigma_i|\sigma\rangle$ ($\sigma_i=\pm1$). 
For this measurement basis, we consider the phase $f_\sigma(t)$ satisfying
\begin{equation}
\frac{df_\sigma(t)}{dt}=-\lambda(t)\left(\sum_{\substack{i,j=1 \\ (i<j)}}^NJ_{ij}\sigma_i\sigma_j+\sum_{i=1}^Nh_i\sigma_i\right).
\label{Eq.fsigma.QA}
\end{equation}
Then, we find that the unitary (\ref{Eq.Vf}) is given by
\begin{equation}
\hat{V}_f(\tau_m)=\exp\left(i\int_0^{\tau_m}dt\ \lambda(t)\hat{H}_P\right),
\label{Eq.Vf.QA}
\end{equation}
and Eq.~(\ref{Eq.ham.phi.u}) is given by
\begin{equation}
\hat{H}_f^V(\tau_m)=\bm{(}1-\lambda(\tau_m)\bm{)}\hat{V}+\hat{H}_\mathrm{LCD}(\tau_m). 
\label{Eq.local.fields}
\end{equation}
The unitary operator (\ref{Eq.Vf.QA}) can be decomposed into the $ZZ$ rotation gates and $Z$ rotation gates without errors due to the commutativity, and the unitary operator with Eq.~(\ref{Eq.local.fields}) can be decomposed into local rotation gates without errors due to its locality. 
Therefore, the time evolution operator can be decomposed into quantum gate operations that have already been realized in experiments without any additional error [errors only arise from Eq.~(\ref{Eq.Uf.Magnus})]. 
More preciusely, the time-evolution operator is given by
\begin{equation}
\begin{aligned}
\hat{U}_f(t_m,t_{m-1})=&\prod_{\substack{i,j=1 \\ (i<j)}}^NR_{ZZ}(a_{ij})\prod_{i=1}^NR_Z(b_i)\prod_{i=1}^NR_X(c_i)\\
&\times\prod_{i=1}^NR_Z(-b_i)\prod_{\substack{i,j=1 \\ (i<j)}}^NR_{ZZ}(-a_{ij})+\mathcal{O}(\delta t_m^3)
\end{aligned}
\end{equation}
where
\begin{equation}
R_W(\bullet_i)=\exp\left(-i\frac{\bullet_i}{2}\hat{W}_i\right),\quad W=X,Y,Z,
\end{equation}
is the single-qubit rotation gate,
\begin{equation}
R_{ZZ}(\bullet_{ij})=\exp\left(-i\frac{\bullet_{ij}}{2}\hat{Z}_i\hat{Z}_j\right),
\end{equation}
is the $ZZ$ rotation gate, and the rotation angles are
\begin{equation}
\begin{aligned}
&a_{ij}=2\int_0^{\tau_m}ds\ \lambda(s)J_{ij},\\
&b_i=2\int_0^{\tau_m}ds\ \lambda(s)h_i-\arctan\frac{\alpha_i(\tau_m)}{\bm{(}1-\lambda(\tau_m)\bm{)}\Gamma_i},\\
&c_i=-2\delta t_m\sqrt{\bm{(}1-\lambda(\tau_m)\bm{)}^2\Gamma_i^2+\alpha_i^2(\tau_m)}.
\end{aligned}
\end{equation}
The number of rotation gates in the total time-evolution operator is $(1/2)N(N+3)M$ because some decomposed factors can be merged.

Note that the choice of measurement bases is not unique. 
For example, we can consider a time-dependent rotated computational basis $|\sigma\rangle=\otimes_{i=1}^N|\sigma_i^{\theta\phi}\rangle$, where $|\sigma_i^{\theta\phi}\rangle=e^{-i(\phi_i/2)\hat{Z}_i}e^{-i(\theta_i/2)\hat{Y}_i}|\sigma_i\rangle$ with time-dependent rotation angles $\phi_i=\phi_i(t)$ and $\theta_i=\theta_i(t)$. 
As long as the rotation angle $\theta_i$ satisfies $\theta_i(T)=0$ (or, $\theta_i(T)=\pm\pi$), it gives the same result as the original quantum annealing. 
%For this measurement basis, we consider the following phase $df_\sigma(t)/dt=\sum_{i=1}^N\beta_i(t)\sigma_i^{\theta\phi}$, where $\sigma_i^{\theta\phi}$ is the eigenvalue of the operator $\hat{Z}_i^{\theta\phi}=e^{-i\frac{\phi_i}{2}\hat{Z}_i}e^{-i\frac{\theta_i}{2}\hat{Y}_i}\hat{Z}_ie^{i\frac{\theta_i}{2}\hat{Y}_i}e^{i\frac{\phi_i}{2}\hat{Z}_i}$, and then 
Then, with appropriate phase, we can separate local terms as $\hat{V}_f(t)$.

%
%==========================================================
%
\section{Comparison with Trotterization}

Straightforward application of the first-order Suzuki-Trotter decomposition gives
\begin{equation}
\begin{aligned}
\hat{U}(t_m,t_{m-1})=&\prod_{\substack{i,j=1 \\ (i<j)}}^NR_{ZZ}(-a_i^\mathrm{ST})\prod_{i=1}^NR_Z(-b_i^\mathrm{ST})\\
&\times\prod_{i=1}^NR_X(-c_i^\mathrm{ST})\prod_{i=1}^NR_Y(d_i^\mathrm{ST})+\mathcal{O}(\delta t_m^2)
\end{aligned}
\end{equation}
and that of the second-order Suzuki-Trotter decomposition gives
\begin{equation}
\begin{aligned}
&\hat{U}(t_m,t_{m-1})=\\
&\prod_{\substack{i,j=1 \\ (i<j)}}^NR_{ZZ}(-a_{ij}^\mathrm{ST}/2)\prod_{i=1}^NR_Z(-b_i^\mathrm{ST}/2)\prod_{i=1}^NR_X(-c_i^\mathrm{ST}/2)\\
&\times\prod_{i=1}^NR_Y(d_i^\mathrm{ST})\prod_{i=1}^NR_X(-c_i^\mathrm{ST}/2)\prod_{i=1}^NR_Z(-b_i^\mathrm{ST}/2)\\
&\times\prod_{\substack{i,j=1 \\ (i<j)}}^NR_{ZZ}(-a_i^\mathrm{ST}/2)+\mathcal{O}(\delta t_m^3). 
\end{aligned}
\end{equation}
where the rotation angles are
\begin{equation}
\begin{aligned}
&a_{ij}^\mathrm{ST}=2\int_{t_m-1}^{t_m}ds\ \lambda(s)J_{ij},\\
&b_i^\mathrm{ST}=2\int_{t_{m-1}}^{t_m}ds\ \lambda(s)h_i,\\
&c_i^\mathrm{ST}=2\int_{t_{m-1}}^{t_m}ds\ \bm{(}1-\lambda(s)\bm{)}\Gamma_i,\\
&d_i^\mathrm{ST}=2\int_{t_{m-1}}^{t_m}ds\ \alpha_i(s).
\end{aligned}
\end{equation}

Some decomposed factors can be merged when we consider the total time-evolution operators. 
As a result, the number of rotation gates becomes $(1/2)N(N+5)M-(1/2)N(N+1)$ for the first-order Suzuki-Trotter decomposition and $(1/2)N(N+7)M$ for the second-order Suzuki-Trotter decomposition. 
Thus, our method can reduce digitization errors and/or the number of gate operations compared with the straightforward applications of the first-order and the second-order Suzuki-Trotter decompositions.

%\begin{widetext}
%\begin{table}
%\begin{tabular}{c|ccc}
%& Our formula & Lie-Trotter formula & 2nd-order Suzuki-Trotter formula \\
%\hline
%Error scaling & $\mathcal{O}(\delta t_m^3)$ & $\mathcal{O}(\delta t_m^2)$ & $\mathcal{O}(\delta t_m^3)$ \\
%Number of rotation gates & $(1/2)N(N+3)M$ & $(1/2)N(N+5)M-(1/2)N(N+1)$ & $(1/2)N(N+7)M$
%\end{tabular}
%\caption{Scaling of errors and the number of rotation gates in our formula, the Lie-Trotter formula, and the second-order Suzuki-Trotter formula. }
%\end{table}
%\end{widetext}

%Note that the order of the decomposed factors is not unique in the Suzuki-Trotter decompositions, and it can affect the number of rotation gates. 
%For example, it becomes $(1/2)N(N+7)M+N$ for the second-order Suzuki-Trotter decomposition if we place the rotation $ZZ$ and $Z$ gates in the middle. 
%Moreover, the order of the decomposed factors affects the amount of errors, while it does not change the scaling of errors. 

\section{\label{Sec.conclusion}Conclusion}

We proposed the decomposition formula which utilized the phase degrees of freedom in the measurement basis. 
The application of the present decomposition formula reduced digitization errors and/or the number of gate operations compared with the straightforward applications of the Lie-Trotter formula and the second-order Suzuki-Trotter formula. 
We note that, by eliminating the Pauli-Y terms with $Z$-axis rotations before the application of the second-order Suzuki-Trotter decomposition, we can acheive the same error scaling and gate efficiency as ours, but our formula gives a new insight into the theory of digitization.

\begin{acknowledgments}
This work was supported by JST Moonshot R\&D Grant Number JPMJMS2061. 
\end{acknowledgments}

% Create the reference section using BibTeX:
\bibliography{DQAbib}

\end{document}